\documentclass[aps,prx,floatfix,twoside,twocolumn,10pt,citeautoscript,superscriptaddress]{revtex4-2}

\usepackage[export]{adjustbox}
\usepackage{amsmath}
\usepackage{amssymb}
\usepackage{booktabs}
\usepackage{comment}
\usepackage{glossaries}
\usepackage{graphicx}
\usepackage[version=4]{mhchem}
\usepackage{siunitx}
\usepackage{tabularx}
\usepackage[colorlinks=true]{hyperref}
\hypersetup{
    pdffitwindow=false,
    pdfstartview={FitH},
    pdfnewwindow=true,
    colorlinks=true,
    linkcolor=blue,
    citecolor=blue,
    filecolor=blue,
    urlcolor=blue,
}

\graphicspath{{figures}}
\hypersetup{pdfauthor={Christopher Linderälv, Nicklas Österbacka, Julia Wiktor, and Paul Erhart}}
\hypersetup{pdftitle={Optical line shapes of color centers in solids from classical autocorrelation functions}}

\setacronymstyle{long-short}
\newacronym{acf}{ACF}{autocorrelation function}
\newacronym{dft}{DFT}{density functional theory}
\newacronym{gf}{GF}{generating function}
\newacronym{md}{MD}{molecular dynamics}
\newacronym{mlp}{MLP}{machine-learning potential}
\newacronym{nep}{NEP}{neuroevolution potential}
\newacronym{pes}{PES}{potential energy surface}
\newacronym{rmse}{RMSE}{root mean square error}
\newacronym{zpl}{ZPL}{zero-phonon line}

\DeclareSIUnit\angstrom{\text{\AA}}
\DeclareSIUnit\Da{\text{Da}}

\renewcommand{\vec}[1]{\boldsymbol{#1}}
\newcommand{\divacancy}{\texorpdfstring{\ce{(V_{Si}V_C)}$_{kk}^0$}{(VSiVC)kk}}

\newcommand{\addOslo}{University of Oslo, Department of Physics, Centre for Material Science and Nanotechnology, P.O. Box 1048, Blindern, Oslo N-0316, Norway}
\newcommand{\addChalmers}{Chalmers University of Technology, Department of Physics, 41296 Gothenburg, Sweden}

\begin{document}

\title{Optical line shapes of color centers in solids from classical autocorrelation functions}

\author{Christopher Linderälv}
\affiliation{\addChalmers}
\affiliation{\addOslo}
\author{Nicklas Österbacka}
\author{Julia Wiktor}
\author{Paul Erhart}
\email{erhart@chalmers.se}
\affiliation{\addChalmers}

\begin{abstract}
Color centers play key roles in applications, including, e.g., solid state lighting and quantum information technology, for which the coupling between their optical and vibrational properties is crucial.
Established methodologies for predicting the optical lineshapes of such emitters rely on the generating function (GF) approach and impose tight constraints on the shape of and relationship between the ground and excited state landscapes, which limits their application range.
Here, we describe an approach based on direct sampling of the underlying auto-correlation functions through molecular dynamics simulations (MD-ACF) that overcomes these restrictions.
The energy landscapes are represented by a machine-learned potential, which provides an accurate yet efficient description of both the ground and excited state landscapes through a \emph{single} model, guaranteeing size-consistent predictions.
We apply this methodology to the \divacancy{} divacancy defect in 4H-SiC, a prototypical color center, which has been studied both experimentally and theoretically.
We demonstrate that at low temperatures the present MD-ACF approach yields predictions in agreement with earlier GF calculations.
Unlike the latter it is, however, also applicable at high temperatures as it is not subject to the same limitations, especially with respect to handling of anharmonicity, and can be applied to study non-crystalline materials.
While we discuss remaining challenges and possible extensions, the methodology presented here already holds the potential to substantially widen the range of computational predictions of the optical properties of color centers and related defects, especially for cases with pronounced anharmonicity and/or large differences between the initial and final states.
\end{abstract}

\maketitle

\section{Introduction}

Color centers in solids are not only the source of the luminous appearance of many gem stones but have also found numerous and widespread technological applications as a platform for harnessing the quantum nature of light.
This includes for example certain types of lasers \cite{Mai60, Gel91}, phosphors for solid state lighting \cite{MckSheLau14, LinKarBet17}], and scintillators for radiation detection \cite{NikYos15, YanTakNak23}.
Furthermore, color centers are emerging as one of the most promising platforms for realizing quantum technologies \cite{WebKoeVar10, Awschalom2018, Wolfowicz2021}, enabling optical initialization and read-out \cite{DohManDel13} as well as single photon emission  \cite{Castelletto_2021, Grosso2017, Aharonovich2016}.
Finally, the sensitivity of the optical properties of color centers to changes in external parameters such as temperature, strain, and electromagnetic fields enables applications in sensing \cite{Gruber97}.
Consequently, the optical line shapes of color centers have become a crucial characteristic, prompting both extensive experimental \cite{TraElbTot16, WigSchPoz19, ShaHasBer20, UdvThiMor20} and theoretical efforts \cite{AlkBucAws14, LinWieErh21, HasLinKra21, JinGovWol21}.

The by far most common methodology for predicting line shapes is based on the \gls{gf} approach \cite{Mar59, MiyDex70} in combination with \gls{dft} calculations.
This method has been shown to yield optical line shapes in good agreement with experiment at low temperatures for various color centers \cite{AlkLyoSte12, AlkBucAws14, PonJiaGia16, ExaHopGro17, JinGovWol21, RazDohMan21, LinWieErh21, LinAbeErh21, BouPonJia21}.
Additional developments to account (partially) for temperature \cite{JinGovWol21}, the Jahn-Teller effect \cite{RazDohMan21}, and the restriction to a one-dimensional configuration coordinate \cite{JiaPonMig19} have further increased the value of this method, and currently it can be considered the reference approach for assessing optical line shapes of color centers in solids.
The \gls{gf} approach, however, involves a number of critical approximations, which can be understood as restrictions on the shapes of and the relation between the ground and excited state \glspl{pes} as well as the treatment of anharmonicity.
As discussed below (\autoref{sect:gf-theory}), these approximations limit the reliability of \gls{gf} predictions at higher temperatures, for anharmonic and/or ionic materials, and in the case of symmetry breaking electronic excitations.

\begin{figure}
    \centering
    \includegraphics[width=0.8\columnwidth]{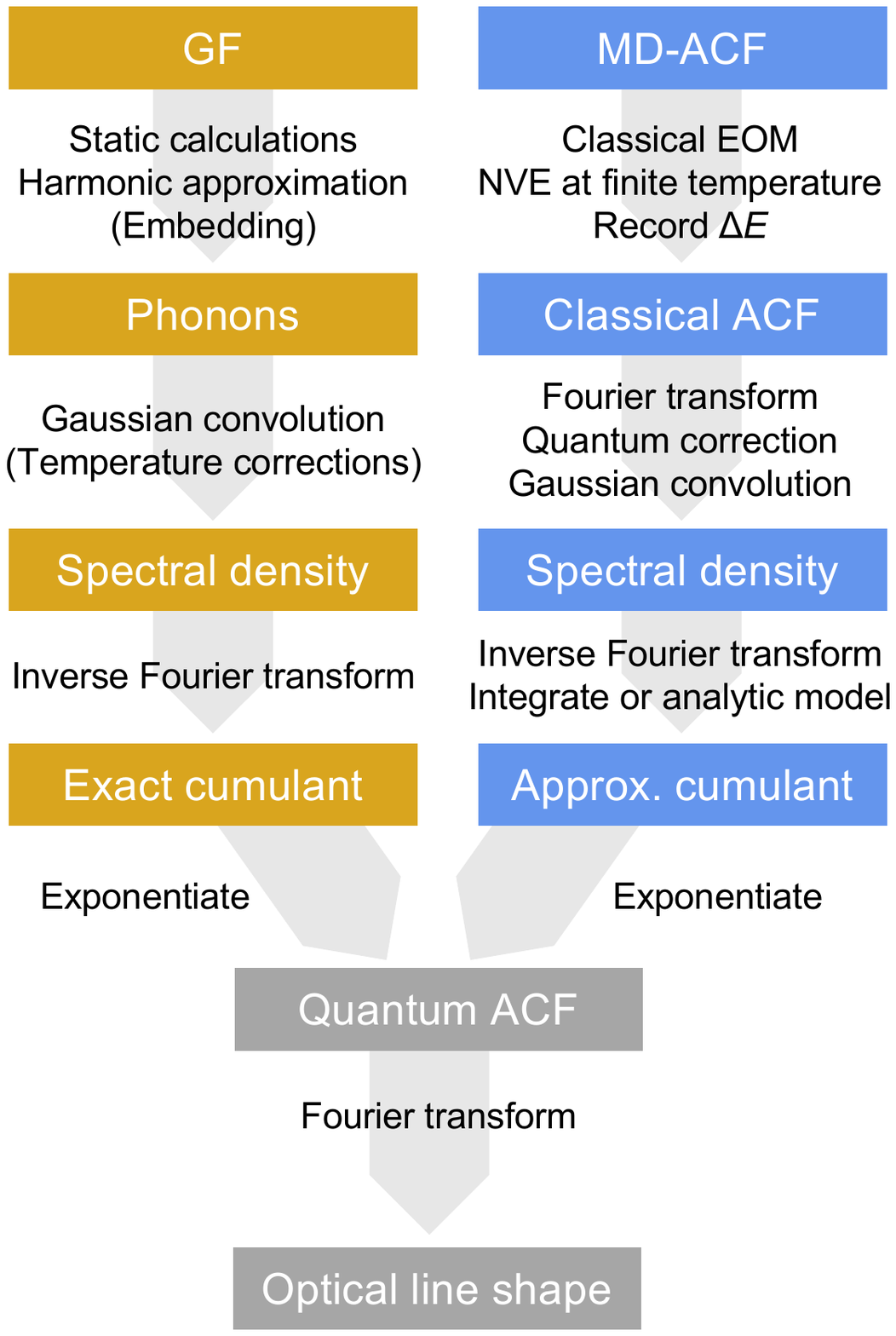}
    \caption{
        \textbf{Workflows.}
        Illustration of the various steps involved in obtaining the optical line shape from the \acrfull{gf} method and the method based on molecular dynamics and autocorrelation function analysis (MD-ACF).
    }
    \label{fig:workflows}
\end{figure}

\begin{figure*}
\centering
\includegraphics[valign=t]{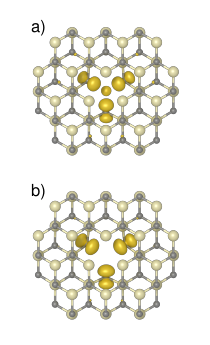}%
\includegraphics[valign=t]{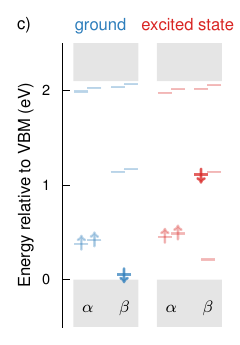}%
\includegraphics[valign=t]{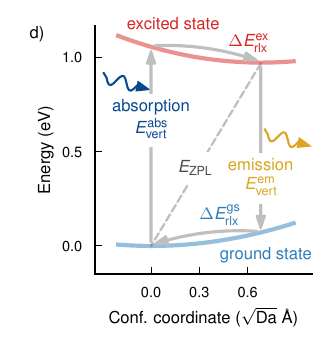}%
\includegraphics[valign=t]{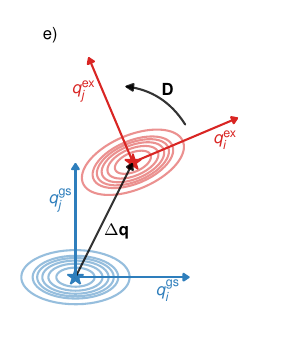}
\caption{
    \textbf{Ground and excited landscapes of the \divacancy{} defect.}
    Defect configuration in the (a) ground and (b) excited state.
    The charge density of the localized electronic states that are occupied in the respective configurations and highlighted in (c) are shown by isosurfaces (also see Figs.~\ref{sfig:defect-orbitals-ground-state} and \ref{sfig:defect-orbitals-excited-state} \cite{sm}).
    (c) Level structure according to the Kohn-Sham eigenstates.
    The localized levels shown in (a,b) are highlighted by bold bars in the $\beta$ spin channel.
    (d) Illustration of the simplified treatment of configuration coordinate in the parallel-mode approximation commonly used in the \gls{gf} approach, which reduces the $3N$-dimensional configuration coordinate to one dimension.
    The minima on the ground and excited state landscapes correspond to the atomic configurations shown in (a) and (b), respectively, and their distance equals $\Delta q$.
    The figure also indicates the vertical excitation energies $E_\text{vert}$, the \gls{zpl} energy $E_\text{ZPL}$ as well as the Stokes shifts $\Delta E_\text{rlx}$.
    (e) Illustration of the more general relationship between the ground and excited state normal modes.
    In this case the normal modes corresponding to ground (gs) and excited state (ex) are translated by a vector $\Delta\vec{q}$ and rotated by the Duschinsky matrix $\mathbf{D}$.
}
\label{fig:defect}
\end{figure*}

Here, we describe a methodology that overcomes several of the most crucial limitations of the \gls{gf} approach, allows one to properly handle the aforementioned cases, and is also applicable to non-crystalline materials.
To this end, we combine \gls{acf} analysis based on \gls{md} simulations with a \gls{mlp} that can handle both the ground and excited state landscapes (\autoref{fig:workflows}).
We demonstrate the utility of this \gls{md}-\gls{acf} scheme by its application to the \divacancy{} divacancy in 4H-SiC (\autoref{fig:defect}).
This prototypical color center has been extensively studied both experimentally \cite{SonCarHas06, ChrFalAnd15, JinGovWol21} and computationally \cite{HasLinKra21, JinGovWol21}, thanks to its potential for applications in quantum technology \cite{ChrFalAnd15, ChrKliDel17, RadWidNie17, WanZhoWan18}.
This defect serves as an ideal benchmark for the present approach, enabling a careful comparison with the \gls{gf} method while providing a challenging test case.

The remainder of this paper is organized as follows.
First, we review the \gls{gf} framework (\autoref{sect:gf-theory}) and the \gls{acf} approach (\autoref{sect:acf-approach}) before introducing a scheme for the construction of a \gls{mlp} that can handle both ground and excited states (\autoref{sect:modeling-energy-landscapes}).
These elements are then combined to analyze the optical fine structure of the aforementioned divacancy defect in 4H-SiC (\autoref{sect:results}).
We provide a detailed discussion of the approximations that are employed and their impact on the predicted line shapes (\autoref{sect:discussion}) before closing with an outlook into possible future developments (\autoref{sect:conclusions-and-outlook}).

\section{Line shape theory}
\label{sect:line-shapes-theory}

First we recap the \gls{gf} approach before describing the \gls{acf} methodology.
The workflows of the two methods are schematically summarized in \autoref{fig:workflows}.

The starting point for either approach for modeling optical line shapes is Fermi's golden rule, from which the following expression for the optical line shape during emission can be derived \cite{BaiBloBar13}
\begin{equation}
    I(\omega)\propto \omega^3 A(\omega),
    \label{eq:fgr}
\end{equation}
where 
\begin{align}
    A(\omega)
    = \int_{-\infty}^{\infty} \text{d}t
    \left\langle \vec{\mu}(0) \vec{\mu}(t) \right\rangle \exp(\mathrm{i}\omega t)
\end{align} 
is the line shape function and
\begin{align}
    \left\langle \vec{\mu}(0) \vec{\mu}(t) \right\rangle = C(t)
    \label{eq:dipole-acf}
\end{align}
is the dipole-dipole \acrfull{acf}.

\subsection{Generating function approach}
\label{sect:gf-theory}

The \gls{gf} approach is based on the (\emph{i}) parallel-mode, (\emph{ii}) Franck-Condon, and (\emph{iii}) harmonic approximations, where we note that the Franck-Condon approximation includes the constant dipole approximation as discussed at the end of this section.
Most of these approximations can be understood as restrictions on the \glspl{pes} (\autoref{fig:defect}d-e).
The harmonic approximation means that both \glspl{pes} are perfectly quadratic.
The parallel-mode approximation further assumes that there is a one-to-one correspondence between modes in the ground and excited state and that one of the \glspl{pes} is simply displaced.
Furthermore, the curvature of the \glspl{pes} is assumed to be exactly the same.
This has to be true for each of the $3N$ vibrational modes. 
Now, these are in general rather strict requirements that can cause qualitative disparity between calculations and experiments.

In the \gls{gf} approach the dipole-dipole \gls{acf} then takes the form \cite{AlkBucAws14}
\begin{align}
    C(t) = \exp(S(t) - S(0)),
\label{eq:corr-gf}
\end{align}
where $S(t)$ is the inverse Fourier transform of the electron-phonon spectral function 
\begin{align}
    S(\omega) = \sum_k s_k \, \delta(\omega - \omega_k).
    \label{eq:electron-phonon-spectral-function}
\end{align}
Here, $s_k$ is the partial Huang-Rhys factor defined as 
\begin{align}
    s_k = \frac{1}{2} \, \omega_k \, \Delta q_k^2,
    \label{eq:partial-hr-factor}
\end{align}
where $\Delta q_k$ is the projection of the ionic configurational difference $\Delta \vec{R}$ on phonon mode $k$ given by
\begin{align}
    \Delta q_k = \sum_a \sqrt{m_a} \, \left<\vec{n}_k^a|\Delta \vec{R}_a\right>.
\end{align}
Here, $a$ runs over all the atoms in the system and $\left|\vec{n}_k\right>$ is the normalized ionic displacement vector corresponding to phonon mode $k$.

The expression for the \gls{acf} in Eq.~\eqref{eq:corr-gf} is obtained under the assumptions that the ground and excited state phonon modes are parallel as well as harmonic, that emission occurs from the vibrational ground state (i.e., low temperature), and that the Franck-Condon approximation holds.
While the low-temperature assumption can to some extent be relaxed by including Bose-Einstein statistics in the correlation function \cite{JinGovWol21}, the other approximations remain difficult to avoid and even control \emph{a priori}.
For example, the evaluation of the partial Huang-Rhys factors according to Eq.~\eqref{eq:partial-hr-factor} strictly requires the phonon modes $k$ in the ground and excited states to have the same frequency and to be linearly related. 

In general, the phonon modes of the ground and excited states are related by a rotation and displacement 
\begin{equation}
    \vec{q}^{\text{ex}}=\mathbf{D} \vec{q}^{\text{gs}} + \Delta \vec{q},
\end{equation}
where $\mathbf{D}$ is the Duschinsky matrix, which introduces a rotation (or mixing) between the phonon modes in the different electronic states and $\Delta \vec{q}$ is the displacement between the minima of the ground and excited state \glspl{pes} (\autoref{fig:defect}e).
In the \gls{gf} approach, the phonon modes are required to be linearly related i.e., ${q}_k^\text{ex}={q}_k^\text{gs} + \Delta q_k$ for some constant $\Delta q_k$.
This approximation can be particularly severe for symmetry breaking excitations/deexcitations and/or in systems with pronounced anharmonicity.

Next, the Franck-Condon approximation is based on the zeroth-order term in a series expansion of the dipole moment and it has been outlined how to include linear terms (Herzberg-Teller factors) \cite{BaiBloBar13}.
Including the linear terms and using different vibrational modes in the initial and final states adds considerable complexity to the task of evaluating the optical line shapes, and has to the best of our knowledge not been implemented and tested for solid state systems.

\subsection{MD-ACF approach}
\label{sect:acf-approach}

In the physical chemistry community modeling line shapes from \glspl{acf} sampled via classical \gls{md} simulations is an established approach, and commonly applied to molecular systems.
While this applies predominantly to vibrational spectra related to infrared and Raman measurements \cite{GasBehMar17, XuRosSch24}, optical spectra can be obtained as well via mixed quantum-classical approaches \cite{Muk82, IslMuk84, ValEisAsp12}.
At the same time, the possibilities to model materials at relevant length and time scales have significantly improved in recent years due to the advent of \glspl{mlp} that are not only accurate and efficient but able to describe systems with increasing chemical, structural, and dynamical complexity.

\subsubsection{Line shape function}

The starting point is again Eq.~\eqref{eq:fgr} and in the Franck-Condon approximation the (quantum) \gls{acf} can be cast in the form of a time-ordered exponential \cite{Muk82}
\begin{align*}
    C(t) = \left\langle \exp\left( \mathrm{i} \int_0^{t} \text{d}t'\, U(t')\right)\right\rangle_\text{ens},
\end{align*}
where $U$ is the \emph{gap} operator, i.e., the energy difference between the initial and final state and the angular brackets indicate an ensemble average.
This exponential can be expanded in cumulants as 
\begin{align*}
    C(t) = \exp\left( \sum\limits_{n=1}^{\infty} g_n(t) \right).
\end{align*}
We define the $m$-th cumulant as 
\begin{align*}
    g_m(t) = \sum\limits_{n=1}^{m} g_n(t).
\end{align*}

For a fully harmonic system $g_m=g_2$ for $m>2$, i.e., $g_2$ describes a harmonic system \emph{exactly} \cite{AndDevHan16}.
The second-order cumulant contains the information present in the two-time classical \gls{acf} of the potential energy difference.
It is given by 
\begin{align}
    g_2(t) = -\int_0^{t} \text{d}t' \int_{0}^{t'} \text{d}t''
    \left\langle \Delta U(0) \, \Delta U(t'') \right\rangle.
    \label{eq:g2}
\end{align}
The cumulant is then related to the optical line shape function as 
\begin{equation} \label{eq:A}
    A(\omega) = \int_{-\infty}^{\infty} \text{d}t
    \exp[\mathrm{i} (\omega - (\underbrace{E_\text{vert} \mp \Delta E_\text{rlx}}_{E_\text{ZPL}})) t]
    \, \exp[-g_2(t)],
\end{equation}
which is centered at the \gls{zpl} $E_\text{ZPL}$, i.e., the energy obtained from the vertical transition energy $E_\text{vert}>0$ and the relaxation energy (or one-sided Stokes shift) $\Delta E_\text{rlx}>0$ where $-$/$+$ are applicable for absorption and emission, respectively (\autoref{fig:defect}d).

In order to obtain the cumulant, we start from the energy difference of the adiabatic \glspl{pes} corresponding to the initial and final electronic states.
The energy difference function can be computed from a classical trajectory propagated with the Hamiltonian corresponding to the initial state.
By propagating on the ground or excited state \gls{pes} we can thus model absorption or emission spectra, respectively.

The classical energy gap operator is 
\begin{align*}
\begin{split}
    \Delta U_\text{cl}(t)
    &= U_\text{cl}^{\text{ex}}(\vec{R}(t)) - U_\text{cl}^\text{gs}(\vec{R}(t)) \\
    &- \left\langle U_\text{cl}^\text{ex}(\vec{R}(t)) - U_\text{cl}^\text{gs}(\vec{R}(t))
    \right\rangle ,
\end{split}
\end{align*}
where $\vec{R}(t)$ is the atomic configuration of the system at time $t$, $\langle \cdot \rangle$ is the average vertical transition energy $\left<E_\text{vert}\right>$, and $U$ is the potential energy of the system.
In this case the energy $\Delta U_\text{cl}$ corresponds to the difference in vibrational energy between the initial and final states.

\subsubsection{Classical ACF}
\label{sect:classical-acf}

The classical \gls{acf} of $\Delta U_\text{cl}$ is the projection of $\Delta U_\text{cl}(t)$ on $\Delta U_\text{cl}(0)$, i.e.,
\begin{align}
    \label{eq:clacf}
    C_\text{cl}(t)
    &= \left<\left<\Delta U_\text{cl}(t) \, \Delta U_\text{cl}(0)\right>_\tau \right>_\text{ens},
\end{align}
where the inner and outer angular brackets indicate an average over time origins and an ensemble average, respectively.

Since $C_\text{cl}(t)$ is symmetric in time, the imaginary part of the classical \gls{acf} is zero, in contrast to the quantum \gls{acf}.
This is related to the phonons in the quantum case carrying a phase as well as the time-energy uncertainty.
The imaginary part of the quantum \gls{acf} can, however, be reconstructed to some extent \emph{a posteriori} from the classical \gls{acf} (see the prefactor in Eq.~\eqref{eq:j} below) \cite{ValEisAsp12}.

Similar to the \gls{gf} approach we apply a broadening to the \gls{acf} to mimic the effect of the instrumental resolution function by convolution with a Gaussian function,
\begin{align}
    \widetilde{C}_\text{cl}(t) = C_\text{cl}(t) \exp(-\gamma t^2).
    \label{eq:gaussian-broadening}
\end{align}
Accordingly we work out the second-order cumulant from $\widetilde{C}_\text{cl}$ instead of the bare $C_\text{cl}$, which smoothens the Fourier transform of the latter, i.e., the spectral density Eq.~\eqref{eq:j}.
This has a similar effect on the spectral density as the approximation of the $\delta$-functions in Eq.~\eqref{eq:electron-phonon-spectral-function} with normalized Gaussians.

\subsubsection{Spectral density function}

The spectral density function $\hat{j}_\text{cl}(\omega)$ is given by \cite{ValEisAsp12}
\begin{align}
    \hat{j}_\text{cl}(\omega) = {\theta(\omega)} \, \frac{1}{\pi} \, f(\beta \omega) \, \widetilde{C}_\text{cl}(\omega),
    \label{eq:j}
\end{align}
where $\theta(\omega)$ is the Heaviside function.
Note that there are several possibilities for choosing the prefactor $f(\beta \omega)$ to include approximate quantum effects \cite{ValEisAsp12, EgoEveSki99} (\autoref{snote:prefactor} \cite{sm}).
In this work we use the harmonic prefactor $\beta\omega/2$, which yields results that are consistent with \gls{gf} calculations \cite{JinGovWol21} and has also been found to be a suitable choice in calculations for the Fenna–Matthews–Olson complex \cite{ValEisAsp12}.
We emphasize that using the harmonic prefactor does \emph{not} imply that we assume the system to behave fully harmonic but rather that the treatment of quantum effects is limited to the harmonic approximation.

The Stokes shift, i.e., the relaxation energy on the final state \gls{pes} (also see \autoref{fig:defect}d) is \cite{LocCup19}
\begin{align}
    \Delta E_\text{rlx} = \int\limits_0^{\infty}\,\text{d}\omega\, \frac{\hat{j}_\text{cl}(\omega)}{\omega}.
    \label{eq:stokes}
\end{align}
Furthermore in the mode-decomposed case the following relation holds \cite{ValEisAsp12}
\begin{align}
    \hat{j}_\text{cl}(\omega)
    = \sum_j \omega_j \, \Delta E_{\text{rlx},j} \, \delta(\omega - \omega_j).
\end{align}
In a fully harmonic system the mode-decomposed Stokes shift can be written in terms of the partial Huang-Rhys factors as $\Delta E_{\text{rlx},j}=\omega_j s_j$.
This yields
\begin{align*}
    \hat{j}_\text{cl}(\omega)
    = \frac{1}{2}\sum_j \omega_j^2 \, s_j \, \delta(\omega - \omega_j),
\end{align*}
which shows that
\begin{align}
    F(\omega) = \hat{j}_\text{cl}/\omega^2
    \label{eq:spectral-density}
\end{align}
plays a similar role as the electron-phonon spectral function $S(\omega)$ in the \gls{gf} approach, see Eq.~\eqref{eq:electron-phonon-spectral-function}.
provides a natural relation between the electron-phonon spectral function in Eq.~\eqref{eq:electron-phonon-spectral-function} and the spectral density computed at finite temperatures.
The relation between the spectral density $F(\omega)$ (or equivalently $\hat{j}_\text{cl}(\omega)$)
and the second-order cumulant $g_2(t)$, see Eq.~\eqref{eq:g2}, is \cite{ZueMonNap19}
\begin{align}
\begin{split}
    g_2(t) =
    - \int_{-\infty}^{\infty} \, &\text{d}\omega\,
   F(\omega)
    \bigg\{
    \coth\left(\frac{\beta \omega}{2} \right) 
    \\
    & \left[\cos(\omega t) - 1\right]
    - \mathrm{i} \left[\sin(\omega t) - \omega t \right]
    \bigg\}.
\end{split}
\label{eq:g2-ex}
\end{align}

The \gls{md}-\gls{acf} approach expressed through Eqs.~\eqref{eq:clacf} to \eqref{eq:g2-ex} relies on the following approximations: (\emph{i}) the truncation of the cumulant expansion at second order, (\emph{ii}) the reconstruction of the quantum time \gls{acf} from the classical \gls{acf}, and (\emph{iii}) the Franck-Condon approximation as discussed in Ref.~\citenum{ZueMonNap19}.

\begin{figure*}
    \centering
    \includegraphics[valign=t]{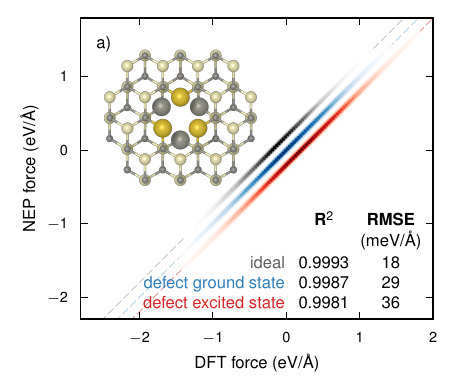}%
    \includegraphics[valign=t]{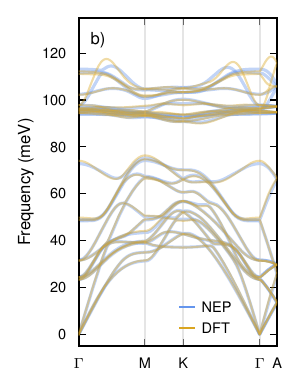}%
    \includegraphics[valign=t]{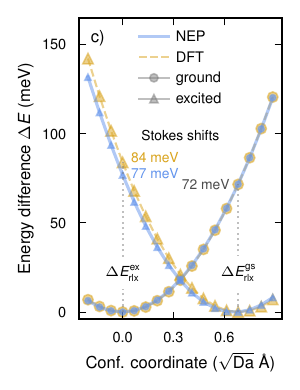}
    \caption{
        \textbf{Machine learning potential model for ground and excited state defect energetics.}
        (a) Parity plots in the form of kernel density estimates for the force components comparing results from the \gls{nep} model with \gls{dft} reference data (also see \autoref{sfig:additional-parity-plots}).
        Coefficients of determination ($R^2$) and \acrfullpl{rmse} for subsets of the data are shown in the table.
        The inset illustrates the defect decoration approach with Si and C atoms associated with the defect shown as large yellow and dark gray atoms.
        (b) Phonon dispersion for the pristine primitive cell from the \gls{nep} model and \gls{dft} calculations (\numproduct{6x6x2} supercell, \num{576} atoms).
        (c) 1D configurational coordinate diagram computed for the ground and excited state of the divacancy (286-atom supercells relaxed with \gls{dft}).
        For ease of comparison the energy difference $\Delta E$ is shown with respect to the minimum of each curve.
        The Stokes shifts $\Delta E_\text{rlx}$ on the excited (ex) and ground state (gs) landscape from \gls{nep} and \gls{dft} calculations are indicated.
    }
    \label{fig:mlp}
\end{figure*}

\section{Modeling energy landscapes}
\label{sect:modeling-energy-landscapes}

\subsection{The \divacancy{} divacancy}

In 4H-SiC Si and C occupy two symmetry inequivalent sites, commonly referred to as $h$ and $k$.
There are thus four different divacancy configurations, corresponding to the combinations $(h,h),(h,k),(k,h)$, and $(k,k)$. 
In this study we consider the \divacancy{} divacancy in 4H-SiC (\autoref{fig:defect}), which hosts a bright transition and has been extensively studied both experimentally and computationally \cite{FalKliBuc14, ChrFalAnd15, CsoIvaSon22}.
This defect features four localized levels in the $\alpha$-spin channel and five localized levels in the $\beta$-spin channel that are occupied by two and one electrons, respectively (\autoref{fig:defect}c; also see Figs. \ref{sfig:defect-orbitals-ground-state} and \ref{sfig:defect-orbitals-excited-state}).
In the ground state the lone electron in the $\beta$-channel occupies the lowermost level while in the first excited state it is promoted to the next higher orbital, which is very nearly degenerate with the next-next higher orbital.

Using \gls{dft} calculations and a 286-atom supercell we obtain a vertical excitation energy for absorption of $E_\text{vert}^\text{abs}=\qty{1.06}{\electronvolt}$ (\autoref{fig:defect}d) and a \gls{zpl} energy of $E_\text{ZPL}=\qty{0.97}{\electronvolt}$ (see \autoref{snote:dft-calculations} for computational details).
This is in line with previous theoretical results predicting the \gls{zpl} at \num{1.0} to \qty{1.1}{\electronvolt} using a comparable level of theory and computational settings \cite{DavIvaArm18, JinGovWol21}.

\subsection{MLP for ground and excited states}

The \gls{md}-\gls{acf} approach outlined in \autoref{sect:acf-approach} relies on a very thorough sampling of both ground and excited \glspl{pes} to ensure convergence of the results.
To this end, one requires simulations on the order of nanoseconds and systems on the order of hundred thousand or more atoms.
As this is far beyond the domain of ab-initio \gls{md} simulations we require an effective yet accurate representation of the \glspl{pes}, which is in principle available via \glspl{mlp}.

While there are a few examples of excited state models for molecular systems \cite{WesMar21, AxeShaGom22}, in condensed matter \glspl{mlp} have been so far primarily employed for the description of ground state \glspl{pes}.
Since we are dealing with an extended system represented by periodic boundary conditions, additional considerations are required.
In particular one must ensure that the formation and excitation energies converge to constant values in the limit of an infinite system.
This prohibits the naive approach of constructing separate \glspl{mlp} for ground and excited \glspl{pes}.
To address this issue, we exploit that the electronic impact of the defect considered here (as well as numerous other defects) is largely limited to its immediate vicinity, as the electronic states of the defect are practically completely localized at the nearest neighbor atoms of the vacancy (\autoref{fig:defect}a,b).
This allows us to construct a \emph{single} model that can handle \emph{both} the ground and the excited state by treating the nearest-neighbor atoms of the divacancy as separate species.
Specifically, we set up a \gls{mlp} that distinguishes bulk Si and C representing atoms not directly involved with the defect as well as \ce{Si_{gs}}, \ce{C_{gs}}, \ce{Si_{ex}}, and \ce{C_{ex}}, which represent the nearest neighbors of the divacancy in the ground (gs) and excited state (ex), respectively.

We adopt the \gls{nep} framework \cite{FanZenZha21, Fan22, FanWanYin22} to construct the \gls{mlp} using the \textsc{gpumd} \cite{FanWanYin22} and \textsc{calorine} packages \cite{LinRahFra24} (see \autoref{snote:construction-of-mlp} for details \cite{FraWikErh23, LarMorBlo17, EriFraErh19, Tog23}).
\Gls{nep} models have proven to be not only accurate but also computationally very efficient, which is an important consideration for the present application.
The reference data includes both pristine (defect-free) and defective structures with varying sizes, with energies and forces from \gls{dft} calculations performed using the Vienna ab-initio simulation package (see \autoref{snote:dft-calculations} for details \cite{Blo94, KreJou99, KreHaf93, KreFur1996-1, KreFur1996-2, PerRuzCso08}).
The model as well as the \gls{dft} reference data used for its construction are available on zenodo \cite{zenodo}.

The final \gls{mlp} accurately reproduces the reference \gls{dft} data for ideal (defect-free) structures as well as defect structures in both the ground and excited state as evident from parity plots as well as the \glspl{rmse} and coefficients of determination $R^2$ (Figs.~\ref{fig:mlp}a and \ref{sfig:additional-parity-plots}).
The model also predicts a phonon dispersion in good with \gls{dft} calculations (\autoref{fig:mlp}b) and the \glspl{pes} along the configuration coordinate closely matches the reference data for both ground and excited states (\autoref{fig:mlp}c).
For the 286-atom cell used for this comparison, we obtain vertical excitation energies of \qty{1.01}{\electronvolt} and \qty{1.06}{\electronvolt} from \gls{nep} and \gls{dft} calculations, respectively, as well as \gls{zpl} energies of \qty{0.94}{\electronvolt} and \qty{0.97}{\electronvolt}, with $\Delta q=\qty{0.68}{\sqrt{\Da}\angstrom}$ in both cases.
This implies that also the Stokes shifts on the ground and excited state landscape are in good agreement (\autoref{fig:mlp}c).

\begin{figure}
\centering
\includegraphics{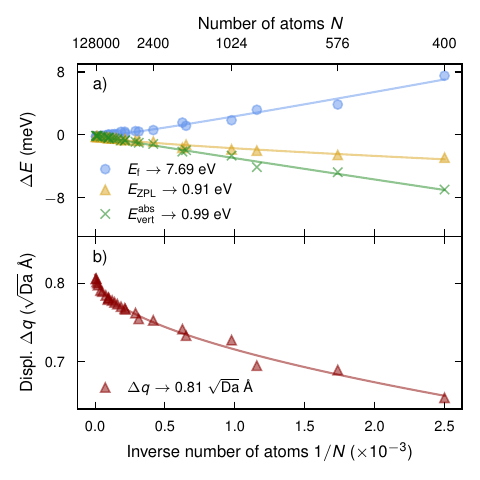}
\caption{
    \textbf{Size convergence of various defect related properties.}
    (a) Variation of the formation energy $E_\text{f}$, \gls{zpl} energy $E_\text{ZPL}$, vertical excitation energy for absorption $E_\text{vert}^\text{abs}$, and (b) the displacement $\Delta q$ with the number of atoms $N$ in the supercell; see \autoref{fig:defect}d for the definition of these quantities.
    Since the leading interaction between defects is due to strain, which scales with the distance $L$ between defects approximately as $1/L^3\sim~1/V^3$, the scaling is shown as a function of the inverse number of atom $1/N$.
}
\label{fig:system-size-scaling}
\end{figure}

The vertical excitation energy for absorption $E_\text{vert}^\text{abs}$, \gls{zpl} energy $E_\text{ZPL}$, and formation energy $E_f$ from the \gls{nep} calculations vary smoothly with system size and converge to constant values as the number of atoms goes to infinity, $N\rightarrow\infty$ (\autoref{fig:system-size-scaling}a).
For system sizes above about \num{2000} atoms the energies deviate by less than \qty{1}{\milli\electronvolt} from their converged values.
The Stokes shifts converge to \qty{71}{\milli\electronvolt} and \qty{80}{\milli\electronvolt} for the ground and excited states, respectively.
We also note that the converged \gls{zpl} energy of \qty{0.91}{\electronvolt} is in good agreement with \gls{dft} calculations using the PBE exchange-correlation functionals for a \num{2400} atom supercell, which yielded a value of \qty{0.94}{\electronvolt} \cite{DavIvaArm18}.

Compared to the energetics the displacement $\Delta q$ exhibits a more pronounced system size dependence, as it increases from \qty{0.68}{\sqrt{\Da}\angstrom} for an 286-atom cell to \qty{0.81}{\sqrt{\Da}\angstrom} in the limit $N\rightarrow\infty$.
This increase can be understood by recalling that even small displacements far from the defect contribute to $\Delta q$ and is testament to the long-ranged (albeit small) elastic strains induced by the defect.

\section{Results}
\label{sect:results}

\subsection{Sampling of the ACF via MD simulations}

Having demonstrated the ability of the \gls{mlp} to describe both the ground and excited state \glspl{pes} on a common footing, we can deploy it in the context of the \gls{md}-\gls{acf} framework.
To this end, we first carried out \gls{md} simulations on both the ground and excited state \glspl{pes} at temperatures of 70, 150, and \qty{300}{\kelvin} using supercells with more \num{1000000} atoms (further computational details are provided in \autoref{snote:md-acf-details}).
These computational parameters were obtained through careful testing.
We note that large system sizes are needed to ensure a dense sampling of the phonon dispersion relative to the Brillouin zone, and also help in reducing the noise in the \gls{acf}.
Thanks to the thin architecture of the \gls{nep} methodology and its efficient implementation on GPUs, the computer time requirements are still modest.

\subsection{Spectral density}

\begin{figure}[b]
\centering
\includegraphics{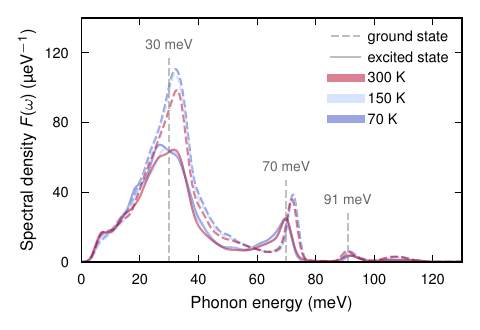}
\caption{
    \textbf{Spectral density.}
    Spectral densities obtained by propagation on the ground and excited state \glspl{pes} at different temperatures using Gaussian broadening with a width of $\gamma=\qty{1.5}{\milli\electronvolt}$, see Eq.~\eqref{eq:gaussian-broadening}.
    The vertical dashed gray lines indicate the center of the main bands in the spectral density for the excited state.
    See \autoref{sfig:spectral-density-broadening-and-prefactor}a,b for an illustration of the effect of broadening.
}
\label{fig:spectral-densities}
\end{figure}

Using Eq.~\eqref{eq:clacf} we computed the classical \gls{acf} from these trajectories, from which the spectral density $F(\omega)$ was obtained via Eqs.~\eqref{eq:j} and \eqref{eq:spectral-density}.
The overall features of the spectral density $F(\omega)$ are consistent for the different temperatures exhibiting a main peak at around \qty{30}{\milli\electronvolt} and a smaller peak at \qty{70}{\milli\electronvolt} (\autoref{fig:spectral-densities}; also see \autoref{sfig:spectral-density-broadening-and-prefactor}a,b).
There are also very minor contributions at higher energies around \qty{110}{\milli\electronvolt}.
It is also apparent that the broadening of the spectra with increasing temperature is readily captured without having to resort to empirical parameters.

The features and the overall shape of the spectral density are consistent with the electron-phonon spectral function $S(\omega)$ in the \gls{gf} approach given by Eq.~\eqref{eq:electron-phonon-spectral-function} \cite{HasLinKra21, JinGovWol21}.
We note that a direct comparison of $S(\omega)$ and $F(\omega)$ would be misleading as the latter is computed at finite temperatures and thus the treatment of phonon occupation factors is relevant (classical vs quantum statistics).

There are few differences between the ground state and excited state spectral density functions.
When propagating on the excited state \gls{pes} the integral over the spectral density $F(\omega)$ at \qty{70}{\kelvin}, as a proxy for the total Huang-Rhys factor, is significantly smaller at \num{1.93}, compared with \num{2.57} for the ground state case (\autoref{fig:spectral-densities}).
These numbers decrease with temperature, with values of \num{1.89}/\num{2.47} at \qty{150}{\kelvin} and \num{1.85}/\num{2.33} at \qty{300}{\kelvin} for the excited/ground state.
Furthermore, the main peak and the peak at around \qty{70}{\milli\electronvolt} are slightly red-shifted in comparison with propagation on the ground state \gls{pes}.

\begin{figure*}[bt]
    \centering
    \includegraphics{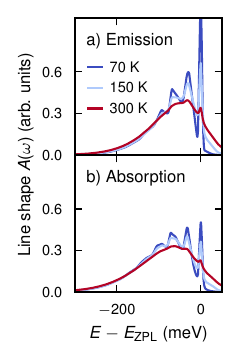}
    \includegraphics{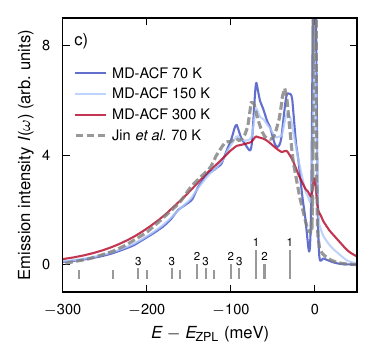}
    \includegraphics{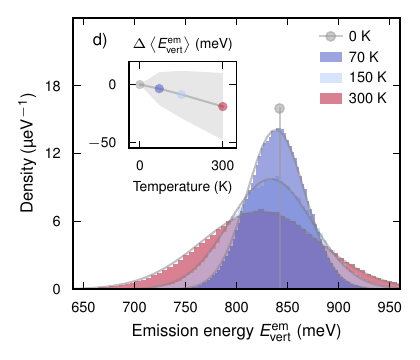}
    \caption{
        \textbf{Line shape functions and emission spectra.}
        (a,b) Line shape functions $A(\omega)$ from Eq.~\eqref{eq:A} obtained by propagation on (a) the excited and (b) ground state \gls{pes} to describe emission and absorption, respectively.
        (c) Emission spectra from the MD-ACF approach described in this work in comparison with calculations using the \gls{gf} approach by Jin \textit{et al.} \cite{JinGovWol21}.
        The vertical gray lines indicate the frequencies that are obtained by combining the bands at \qty{30}{\milli\electronvolt} and \qty{70}{\milli\electronvolt} observable in the spectral density (\autoref{fig:spectral-densities}).
        The order of the combination is given by the numbers and indicated by the length of the lines.
        (d) Distribution of vertical transition energies for propagation on the excited state \gls{pes}.
        The distributions are rather well described by Gaussians (solid gray lines), indicating that truncating the cumulant expansion at second order is a good approximation in this case.
        The inset shows the shift of the average emission energy $\left< E_\text{vert}^\text{em}\right>$ with respect to the emission energy at \qty{0}{\kelvin}.
    }
    \label{fig:spectra}
\end{figure*}

Finally we note that the feature at \qty{70}{\milli\electronvolt} (i.e., before application of Gaussian broadening) is fundamentally actually very narrow ($\lesssim\qty{1}{\milli\electronvolt}$; \autoref{sfig:spectral-density-broadening-and-prefactor}a,b) whereas the feature at \qty{30}{\milli\electronvolt} has an intrinsic full width-half maximum of about \qty{14}{\milli\electronvolt}.
This suggests that the latter is the result of the coupling of many modes whereas the former originates from only one or at most few modes.
The difference is, however, obscured by the Gaussian smearing, which is applied here to mimic instrumental broadening present in experimental measurements.

\subsection{Emission line shape}

The emission line shapes $A(\omega)$, see Eq.~\eqref{eq:A}, exhibit a series of sharp features that broaden with increasing temperature (\autoref{fig:spectra}a).
The spacing between these features can be understood as resulting from the combination of the bands observable in the spectral density with lower order combinations giving stronger features (\autoref{fig:spectral-densities}).
This becomes even more apparent when considering the emission intensity $I(\omega$), see Eq.~\eqref{eq:fgr} (\autoref{fig:spectra}c).
The two most prominent side bands at $\widetilde{\omega}_1=\qty{30}{\milli\electronvolt}$ and $\widetilde{\omega}_2=\qty{70}{\milli\electronvolt}$ can thus be identified as first order excitations of the respective bands, while second order excitations give rise to the shoulders at \qty{60}{\milli\electronvolt} ($2\times\widetilde{\omega}_1$) and \qty{140}{\milli\electronvolt} ($2\times\widetilde{\omega}_2$) as well as the peak at \qty{100}{\milli\electronvolt} ($\widetilde{\omega}_1+\widetilde{\omega}_2$).

The features are most pronounced in the spectrum at \qty{70}{\kelvin}, in accordance with published line shapes at \qty{0}{\kelvin} \cite{HasLinKra21} and especially the \qty{70}{\kelvin} spectrum computed using a temperature-extended \gls{gf} approach with force constants embedding \cite{JinGovWol21}.
Overall, the line shape is rather similar, both the intensity of the peaks and the low energy tail.
The small remaining deviations can likely be attributed to differences in the computational parameters, in particular the exchange-correlation functionals as the present work used PBEsol while the DDE functional \cite{SkoGovGal14} was employed in Ref.~\citenum{JinGovWol21}.
Notably the \gls{md}-\gls{acf} results capture the significant broadening of the spectrum at \qty{300}{\kelvin} as well as the reduction in the intensity of the \gls{zpl}.

One of the main approximations of the \gls{md}-\gls{acf} method is the assumption that the cumulant expansion can be truncated at second order. 
The reliability of this approximation can be checked by investigating the distribution of the emission energies observed during an \gls{md} simulation.
While one does notice that that the distribution becomes slightly asymmetric at \qty{300}{\kelvin} (\autoref{fig:spectra}d), the distributions are still very well represented by Gaussians and hence the truncation at second order is justified in the present case.
Our simulations also allow us to obtain the temperature dependence of the average emission energy $\left<E_\text{vert}^\text{em}\right>$ (i.e., the mean of the distributions in \autoref{fig:spectra}d), which we find to red-shift by \qty{19}{\milli\electronvolt} between \num{0} and \qty{300}{\kelvin} (\autoref{fig:spectra}d, inset).

Here, we do not report absorption spectra due to the degeneracy of the excited states.
The latter causes electronic and vibrational states to mix (which is not included in the current model), which in turn prevents a meaningful with experiment.
For reference we do, however, include a comparison of the spectral densities from propagation on the ground and excited states (\autoref{fig:spectra}a,b).

\section{Discussion}
\label{sect:discussion}

\subsection{Anharmonicity and beyond the parallel-mode approximation}
\label{sect:discussion-anharmonicity}

A crucial parameter for any color center is the (total) Huang-Rhys factor.
It can be obtained from the displacement $\Delta q$ via
\begin{align}
    S = \omega_\text{eff} \, \Delta q^2 / 2.
\end{align}
where $\omega_\text{eff}$ is an effective frequency representing the curvature at the minimum of the \gls{pes}, or equivalently by integrating over the spectral function $S(\omega)$, see Eq.~\eqref{eq:electron-phonon-spectral-function}.
The Huang-Rhys factor has a profound effect on the optical line shape.
Smaller values ($S\lesssim 2$) commonly give rise to structured line shapes with distinct peaks whereas large values give rise to wide Gaussian side bands \cite{MiyDex70}.

For applications in lighting, color centers are used to red-shift the energy of absorbed light \cite{BlaGra94}, which requires moderate to large Huang-Rhys factors.
By contrast for applications in quantum information theory, one typically aims for a small to moderate Huang-Rhys factors in order to maintain coherent emission.
As the displacement between the ground and excited \glspl{pes} increases (and thus the Huang-Rhys factor), vertical transitions are increasingly likely to terminate in the anharmonic region of the receiving \gls{pes}.
Simultaneously the parallel-mode approximation and the restriction to a one-dimensional configuration coordinate (\autoref{sect:gf-theory} and \autoref{fig:defect}d) become increasingly questionable.
As the \gls{md}-\gls{acf} approach (\autoref{sect:acf-approach}) in combination with a model for the ground and excited state landscapes (\autoref{sect:modeling-energy-landscapes}) can sample the configuration space over a very wide region it is not bound by these approximations and particularly well suited for such cases.
Here, we have chosen a defect with an intermediate Huang-Rhys factor of 2.5 to 2.8 at \qty{0}{\kelvin} \cite{JinGovWol21} as it provides a good reference point to demonstrate and benchmark the \gls{md}-\gls{acf} approach.
We emphasize, however, that the latter should be even more powerful when applied to systems with larger Huang-Rhys factors, stronger anharmonicity and/or large differences in the vibrational structure of ground and excited state.

In the limit of exactly parallel modes the line shape function $A(\omega)$ is symmetric around $\omega=0$.
For the \divacancy{} defect considered here, the differences in the line shape functions $A(\omega)$ obtained by propagation on the ground and excited state \glspl{pes} (\autoref{fig:spectra}) reveal, however, an asymmetry and thus the limitation of the parallel-mode approximation even in this relatively simple case.
The asymmetry is also apparent in the Huang-Rhys factor, which can be approximately measured by the integral of $F(\omega)$, which equals \num{1.93} for emission and \num{2.57} for absorption at \qty{70}{\kelvin}.
While these numbers cannot be compared \emph{directly} with the \gls{gf} approach (see comment following Eq.~\eqref{eq:spectral-density}), we note that they are comparable to the \qty{0}{\kelvin} Huang-Rhys factor of 2.5 to 2.8 obtained in Ref.~\citenum{JinGovWol21}.
The latter reference also analyzed the limitations of the parallel-mode approximation using a one-dimensional model with different effective frequencies for ground and excited states.
It was concluded that the difference of around \qty{5}{\milli\electronvolt} resulted in minor errors at lower temperatures but in non-negligible errors at high temperatures.
While this points into a similar direction as the present analysis, our results suggest that differences might already be noticeable at lower temperatures.

The spectral densities obtained by propagation on the ground and excited landscapes exhibit some notable differences (\autoref{fig:spectra}a,b), which implies that the optical line shapes are also different.
In terms of the efficiency of the sampling we note that it is still unclear which \gls{pes} should be chosen for propagating the system \emph{in general} \cite{Muk82}.
In this context, the expressions for the cumulant for propagating the system on the final or even an average of the initial and final \gls{pes} for the emission process has been provided in the literature \cite{Muk82}. 
The choice of \gls{pes} to propagate on may ultimately be system dependent and a matter of sampling.
For this system, however, propagation on the initial state \gls{pes} provides a very good agreement with other theoretical work and measurements on the emission line shape \cite{JinGovWol21}.

\subsection{Modeling ground and excited state landscapes}
\label{sect:discussion-mlp}

We have introduced a simple yet effective procedure for constructing \glspl{mlp} that can accurately handle ground and excited states in a computationally efficient manner (\autoref{sect:modeling-energy-landscapes}).
It can be straightforwardly extended to include further excited states by adding corresponding ``marker'' species.
The approach is generally applicable to defects and related excitations such as self-trapped excitons and polarons, as long as they do not diffuse over the relevant time scales at the temperatures of interest.
This category includes a huge number of cases, opening the possibility to quantitatively study the opto-vibrational coupling in these systems at elevated temperatures, a challenge that dates back to the early work by Born and Huang \cite{BorHua54}.

Finally, we note that thanks to the general form of the \gls{mlp} it allows to also readily investigate not only the effect of temperature but also strain.

\section{Conclusions and outlook}
\label{sect:conclusions-and-outlook}

In this work we have described an approach for predicting the optical spectra of defects that overcomes the restrictions on the ground and excited landscapes common in calculations based on the \gls{gf} approach (\autoref{sect:gf-theory}) and is also applicable to non-crystalline materials.
In this approach the \gls{acf} of the gap operator is sampled through \gls{md} simulations (\autoref{sect:acf-approach}) using a \gls{mlp} based on the \gls{nep} framework to model the ground and excited state landscapes (\autoref{sect:modeling-energy-landscapes}).
We demonstrated a simple yet effective approach for constructing \glspl{mlp} that can accurately yet efficiently describe both the ground and excited state landscapes through a \emph{single} model, guaranteeing size-consistent predictions.
In this context the efficiency of the \gls{nep} framework and its implementation in the \textsc{gpumd} package have been beneficial as they allowed us to sample the \gls{acf} at a moderate computational cost.

To demonstrate this methodology we considered the \divacancy{} divacancy in 4H-SiC, which has been previously investigated both experimentally and theoretically (\autoref{fig:defect}).
This defect provides a suitable challenge since it features a moderate Huang-Rhys factor, such that the approximations commonly required in the \gls{gf} approach are warranted at low temperature but become more severe with increasing temperature \cite{JinGovWol21}.
The \gls{md}-\gls{acf} approach predicts low-temperature spectra that are in good agreement with earlier \gls{gf} calculations (\autoref{fig:spectra}c).
At the same time the broadening of the spectra with increasing temperature is readily captured without having to resort to empirical parameters.
These results demonstrate the reliability and potential of the \gls{md}-\gls{acf} approach.
As the latter is not limited by constraints on the relation between the initial and final state landscapes commonly required in the \gls{gf} approach, it should be particularly suitable for studying systems at higher temperatures, with pronounced anharmonicity, and/or large differences between the initial and final landscapes (\autoref{sect:discussion-anharmonicity}).

The \gls{md}-\gls{acf} method applied to color centers in solids is thus an important step towards a more comprehensive description of the coupling between radiative transitions and vibrational degrees of freedom in condensed matter systems.
We envision that this method can be extended to capture additional elusive effects.
This includes, e.g., the commonly used assumption of a constant transition dipole moment, as the latter can be represented through a machine learning model using for example the tensorial \gls{nep} formalism \cite{XuRosSch24}.
This would allow one to sample the dipole-dipole \gls{acf}, see Eq.~\eqref{eq:dipole-acf}, directly as commonly done when predicting infrared spectra of molecules and liquids.
Moreover the dissipative dynamics that leads to \gls{zpl} broadening may be possible to address by including higher order cumulants \cite{AndDevHan16}.
This would require additional algorithmic development since already three-time correlation functions are prohibitively expensive to calculate directly for long time series.
In this context, we note that it might also be possible to approximate the quantum \gls{acf} and bypass the phenomenological prefactor via path integral molecular dynamics in the form of, e.g., ring polymer or Matsubara dynamics \cite{HelWilMuo15}. 
Finally, we recall that the current approach for constructing models for ground and excited \glspl{pes} is limited to (the large class of) immobile (non-diffusing) defects, although it does indicate a potential pathway for generalizing the construction of \gls{mlp} models that are not subject to this constraint (\autoref{sect:discussion-mlp}).

\section*{Data availability}

The \gls{nep} model as well as the database of \gls{dft} calculations used to train this model are available on zenodo at \url{https://doi.org/10.5281/zenodo.13284738}.

\section*{Acknowledgments}

We gratefully acknowledge funding from the Swedish Research Council (Nos.~2020-04935 and 2021-05072) as well as computational resources provided by the National Academic Infrastructure for Supercomputing in Sweden at NSC, PDC, and C3SE partially funded by the Swedish Research Council through grant agreement No.~2022-06725.
Parts of the computations were performed on resources provided by UNINETT Sigma2 --- the National Infrastructure for High Performance Computing and Data Storage in Norway. 
C.~L. acknowledges support provided by the Research Council of Norway and the University of Oslo through the research project QuTe (no. 325573, FriPro ToppForsk-program).
J.~W. acknowledges funding from the Swedish Strategic Research Foundation through a Future Research Leader programme (FFL21-0129).

\end{document}